\documentclass{webofc}
\usepackage[varg]{txfonts}   
\begin{document}
\title{Medium modification of strange hadronic resonances  \\ at SIS, RHIC and LHC energies}

\author{\firstname{Laura} \lastname{Tolos}\inst{1,2}\fnsep\thanks{\email{tolos@ice.csic.es}}         
}

\institute{Institute of Space Sciences (CSIC-IEEC), Campus Universitat Aut\`onoma de Barcelona, \\ Carrer de Can Magrans, s/n, 08193 Cerdanyola del Vall\`es,
Spain 
\and
Frankfurt Institute for Advanced Studies, Goethe University Frankfurt, \\ Ruth-Moufang-Str. 1,
60438 Frankfurt am Main, Germany}

\abstract{%
The properties of strange pseudoscalar and vectors mesons as well as strange baryon resonances in dense matter are reviewed. Some open questions on the  properties of strange hadrons in medium are addressed, such as the experimental signatures of in-medium effects coming from the hadronic phase on the final observables in heavy-ion collisions  for the experimental conditions at SIS, RHIC and LHC energies.
}
\maketitle
\section{Introduction}
\label{intro}
Strangeness production and propagation is one of the major research domains in heavy-ion collisions (HICs) for energies ranging from SIS/GSI, RHIC/BNL and LHC/CERN  up to the forthcoming CBM/FAIR, BMN/NICA, BESII/RHIC and J-PARC-HI.  In particular, understanding the dynamics  of strange mesons, such as pseudoscalar ($K$,$\bar K$) and vector ($K^*$,$\bar K^*$) mesons, as well as strange baryon resonances in HICs is still a challenge \cite{Hartnack:2011cn}.

The production of $K$ and $\bar K$ close to the threshold energy has been thoroughly investigated in low-energy HICs by the KaoS \cite{Forster:2007qk}, FOPI \cite{fopi} and HADES Collaborations \cite{hades1}. The analysis of experimental data together with microscopic transport approaches have permitted drawing several conclusions regarding the production mechanisms and the freeze-out conditions of strange mesons. Still, a simultaneous description of all observables involving $K$ and $\bar K$ production is missing.  

Recently HADES Collaboration has also reported a deep sub-threshold $K^{*0}$ production in Ar+KCl collisions \cite{hades2}. The STAR Collaboration has addressed the $K^{*0}$ meson production in  Au+Au, Cu+Cu, d+Au and p+p at high-energy HICs of $\sqrt{s_{NN}}=$62.4 and 200 GeV  \cite{star}, while attenuation of the $K^{*0}$ and $\bar K^{*0}$ states in the hadronic phase of the expanding fireball in HICs has been observed by the NA49 Collaboration \cite{NA49}. Moreover, the production of $K^{*0}$ in p+p, p+Pb and Pb+Pb collisions has been studied by the ALICE Collaboration, also observing a centrality-dependent and multiplicity-dependent suppression of the $K^{*0}/K$ ratio \cite{badala}.

All such findings indicate that a reliable determination of the interaction of strange mesons with the surrounding hadronic environment is important, since it might help to give answer to several open questions in matter, such as the experimental signatures of in-medium effects coming from the hadronic phase on the final observables in HICs. Therefore, in this paper  the properties of strange pseudoscalar and vector mesons and some strange baryon resonances in matter are reviewed for the experimental conditions ranging from SIS to RHIC and LHC energies.

\section{Strange pseudoscalar mesons}
\label{pseudo}

The properties of the  strange pseudoscalar mesons ($K$,$\bar K$) in hot dense matter are described by means of their self-energies and, hence, spectral functions. The $K$ and $\bar K$ self-energies  are obtained from the $s$- and $p$-waves in-medium $KN$ and $\bar KN$
interactions within a chiral unitary approach \cite{Tolos:2006ny,Tolos:2008di}. On the one hand, the $s$-wave amplitude of the $\bar K N$ comes from the Weinberg-Tomozawa term of the chiral Lagrangian. Unitarization in coupled channels is imposed on the on-shell amplitudes with a cutoff regularization. The $\Lambda(1405)$ resonance in the $I=0$ channel is generated dynamically, whereas a satisfactory description of low-energy scattering observables is achieved. On the other hand, the $s$-wave $K N$ effective interaction is also obtained from unitarization using the same cutoff parameter. 

The in-medium $s$-wave amplitude accounts for Pauli-blocking effects, mean-field binding on the nucleons and hyperons via a $\sigma-\omega$ model, and the dressing of the $\pi$ and $K$/$\bar K$ propagators. The self-energy is then obtained in a self-consistent manner summing the in-medium on-shell amplitudes for the different isospins over the nucleon Fermi distribution at a given temperature \cite{Tolos:2008di}. In the case of  the  $\bar K$ meson, the model includes, in addition, a $p$-wave contribution to the self-energy from hyperon-hole ($Yh$)
excitations, where $Y$ stands for $\Lambda$, $\Sigma$ and
$\Sigma^*$ states. For the $K$ meson the $p$-wave self-energy results from
$YN^{-1}$ excitations in crossed kinematics. 

\begin{figure}[t]
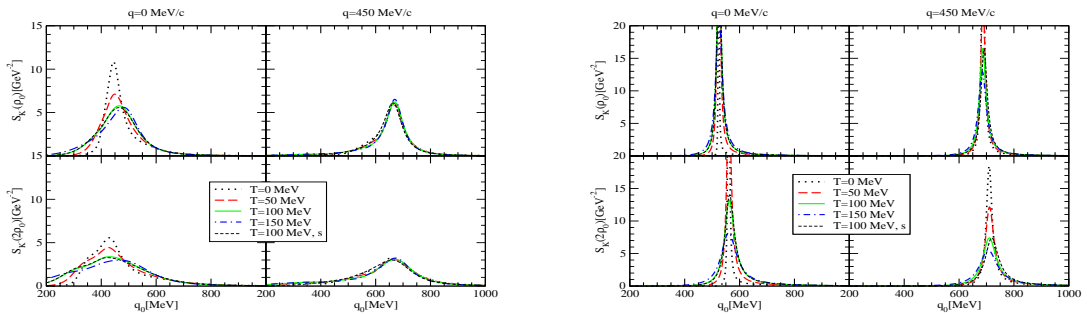

\begin{center}
\includegraphics[height=4 cm, width=6.5cm]{fig6.eps}
\hfill
\includegraphics[height=4 cm, width=6.5cm]{fig7.eps}
\caption{ $\bar K$ (left) and $K$ (right) spectral functions in hot dense matter. Both plots are taken from \cite{Tolos:2008di}. }
 \label{fig1}
\end{center}
\end{figure}

Once the self-energies are determined, the corresponding
$K$ and $\bar K$ spectral functions are obtained, as shown in Fig.~\ref{fig1}.  The $\bar K$ spectral function
(left) shows a broad peak resulting from a strong mixing between the
quasi-particle peak and the $\Lambda(1405)N^{-1}$ and
$YN^{-1}$ excitations. 
These $p$-wave $YN^{-1}$ subthreshold excitations affect mainly the properties of
the ${\bar K}$ at finite momentum. Temperature and density soften the $p$-wave
contributions to the spectral function at the quasi-particle energy. As for the
$K$ meson, the spectral function (right) shows a narrow
quasi-particle peak which dilutes with temperature and density as the phase
space for $KN$ states increases, while shifting it to higher energies with increasing density due to the repulsive character of the $KN$ interaction.

\begin{figure}[t]
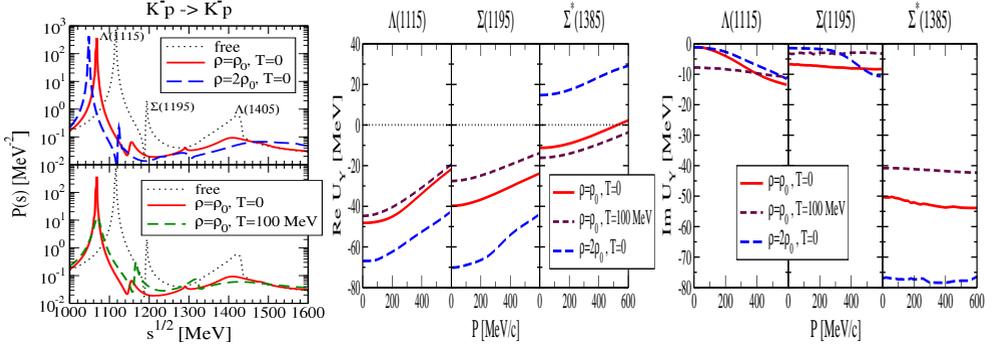

\centering
\includegraphics[height=4.5cm,width=0.3\textwidth,clip]{fig1a.eps} 
\includegraphics[height=4.5cm,width=0.3\textwidth,clip]{ReU-hyp.eps} 
\includegraphics[height=4.5cm,width=0.3\textwidth,clip]{ImU-hyp.eps} 
\caption{Left: In-medium transition probability ${\cal P}$ at zero total three-momentum of the meson-baryon pair for $K^-p$. Middle and Right: The real and imaginary part of the hyperon potentials. All plots are taken from \cite{Cabrera:2014lca}.}
\label{fig:kmp-vs-kmn}       
\end{figure}

Moreover, the in-medium amplitudes in terms of the transition probabilities give access to the behaviour of $\Lambda(1115)$, $\Lambda(1405)$, $\Sigma(1195)$ and $\Sigma^*(1385)$ in matter. As an example, in Fig.~\ref{fig:kmp-vs-kmn}  the transition probability for the $K^-p$ elastic reactions is shown as a function of the meson-baryon center-of-mass energy at total vanishing three-momentum. The $K^-p$ state is an admixture of isospin $I=0,1$ and, thus, the two $\Lambda$ states and the $\Sigma$ show up in the spectrum, as seen in  the left panel of Fig.~\ref{fig:kmp-vs-kmn}.  In the medium,  the structure of the $\Lambda(1405)$ is washed out at normal matter density, whereas the $p$-wave ground states experience moderately attractive mass shifts as density is increased. Temperature  broadens the $\Lambda(1115)$ and $\Sigma(1195)$ states as compared to the vacuum case.

In a more quantitative manner, from the analysis of the transition probabilities one can extract information on the single-particle  potentials of $\Lambda(1115)$, $\Sigma(1195)$ and $\Sigma^*(1385)$ hyperons at finite momentum, density and temperature (middle and right panels of Fig.~\ref{fig:kmp-vs-kmn}). At finite density  the $\Lambda$ and the $\Sigma$ show a potential of roughly -50 and -40~MeV, respectively, at normal matter density $\rho_0$ and zero temperature. Both hyperons acquire a finite decay width with density and temperature, reflecting the probability to be absorbed by the nuclear medium or have quasi-elastic scattering processes. The $\Sigma^*$ develops an attractive potential of about -10~MeV at $\rho_0$ and zero temperature, that turns into a small repulsion for increasing densities, while its decay width is notably enhanced at finite density due to the opening of new absorption channels as pions are dressed. The effect of the temperature in this case is moderate due to the important phase space already available at zero temperature. Also, one finds that the $\Lambda$, $\Sigma$ and $\Sigma^*$ potentials have a smooth behavior with momentum \cite{Cabrera:2014lca}.

At this stage one can address the importance of the in-medium properties of strange hadrons in their production and production in HICs. A systematic study of the experimental results of KaoS collaboration together with a detailed comparison to transport model calculations indicated that, among others, the $K^+$ feel repulsion while the $K^-$ attraction \cite{Forster:2007qk}. However, the conclusions of Leifels and collaborators in SQM2017 conference \cite{leifels} on recent results from HADES and FOPI indicate that, while the $K^+$ shows a repulsive interaction in matter, the $\Phi$ decay into $K^-$  washes out the effects of the $K^-$ potential in the spectra and flow. Thus, more systematic and high statistic data on $K^-$ production are necessary, while further information in elementary reactions is desirable.  The question still remains on the importance of the in-medium effects on the production and propagation of strangeness in HICs. 

\section{Strange vector mesons}
\label{vector}

\begin{figure}[t]
\begin{center}
\includegraphics[width=0.45\textwidth,height=3.5cm]{spectral_ksn.eps}
\hfill
\includegraphics[width=0.5\textwidth,height=3.5cm]{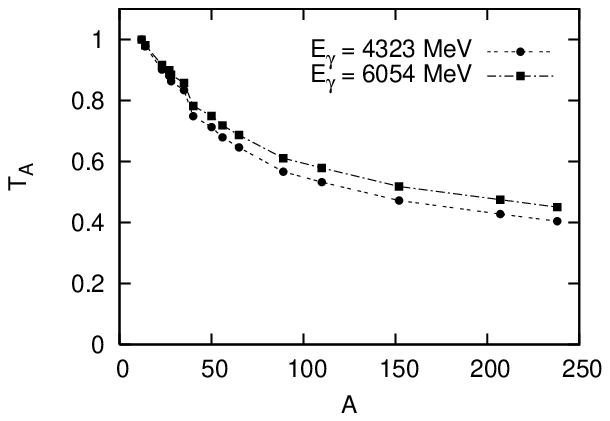}
\caption{ Left: The $\bar K^*$ spectral function at  zero momentum for different  densities. Right: Transparency ratio for  $\gamma A \to K^+ K^{*-} A'$. Both plots are taken from \cite{tolos10}.}
\label{fig:spec-trans}
\end{center}
\end{figure}

In this section results from different groups on the $K^*$ and $\bar K^*$  interactions in free space and in matter are presented as well as some recent conclusions on their production and propagation in low- and high-energy HICs.

The  $\bar K^*$  self-energy in nuclear matter and, hence, spectral function is obtained  within the
hidden gauge formalism, taking into account  two sources for the
modification of the $s$-wave $\bar K^*$  self-energy in nuclear matter \cite{tolos10}. On the one hand, the
contribution associated to the decay mode $\bar K \pi$ modified by nuclear
medium effects on the $\bar K$ and $\pi$ mesons,  which accounts for
the $\bar K^* N \to \bar K N, \pi Y, \bar K \pi N, \pi \pi Y \dots$ processes, 
with $Y=\Lambda,\Sigma$. And, on the other hand, the contribution associated to the interaction
of the $\bar K^*$ with the nucleons in the medium, which considers the  $\bar K^* N \to \bar K^* N$ process, as well as channels involving other vector mesons, such as $\bar K^* N\to \rho Y, \omega Y, \phi Y,
\dots$.  As a result, the interaction of $\bar K^*N$ and coupled channels in free space gives rise to the appearance of two resonances, $\Lambda(1783)$ and $\Sigma(1830)$, which can be identified with the observed $\Lambda(1800)$ and $\Sigma(1750)$, respectively \cite{Oset:2009vf}. These states have been also generated dynamically  in subsequent works on $\bar K^*N$ interaction that incorporate pseudoscalar meson-baryon states as well as the use of an interaction kernel beyond t-channel \cite{Khemchandani:2011et}.

The $\bar K^*$ meson spectral function is displayed in the left panel of
Fig.~\ref{fig:spec-trans} as a function of the meson energy $q_0$, for zero
momentum and different densities. The dashed line refers to
the calculation in free space, where only the $\bar K \pi$ decay channel
contributes, while the other three lines correspond to the self-consistent
calculations,  which incorporate $\bar K^* \rightarrow \bar K
\pi$ in the medium, as well as the quasielastic $\bar K^* N \to \bar K^* N$ and
other $\bar K^* N\to V B$ processes, with $V$ and $B$ denoting vectors and baryons, respectively.  The structures above the quasiparticle
peak correspond to the dynamically generated $\Lambda(1783) N^{-1}$ and
$\Sigma(1830) N^{-1}$ excitations. Density effects result in a dilution and
merging of those resonant-hole states, together with a  broadening of the
spectral function due to the increase of collisional and absorption processes.
As a result, the $\bar K^*$ spectral function spreads substantially in the medium, becoming 
five times bigger at $\rho_0$ than in free space.

In order to test experimentally the broadening of the $\bar K^*$ spectral function, one can study the nuclear transparency ratio. The idea is to compare the cross sections of the photoproduction reaction $\gamma A \to K^+ K^{*-} A'$ in
different nuclei, and trace them to the in medium $K^{*-}$ width. In the right panel of Fig. \ref{fig:spec-trans}  the transparency ratio is shown for
different nuclei for two different
energies of the photon in the lab frame of $4.3$ GeV and
$6$ GeV.  A very strong attenuation of the $\bar{K}^*$
survival probability is found due to the decay  $\bar{K}^*\to
\bar{K}\pi$ or absorption channels 
$\bar{K}^*N\to \bar K N, \pi Y, \bar K \pi N, \pi \pi Y, \bar K^* N, \rho Y,
\omega Y, \phi Y, \dots$  with increasing nuclear-mass number $A$. This is due
to the larger path that the $\bar{K}^*$ has to follow before it leaves the
nucleus, having then more chances to decay or get absorbed.

\begin{figure}[t]
\includegraphics[height=4cm, width=0.4\textwidth]{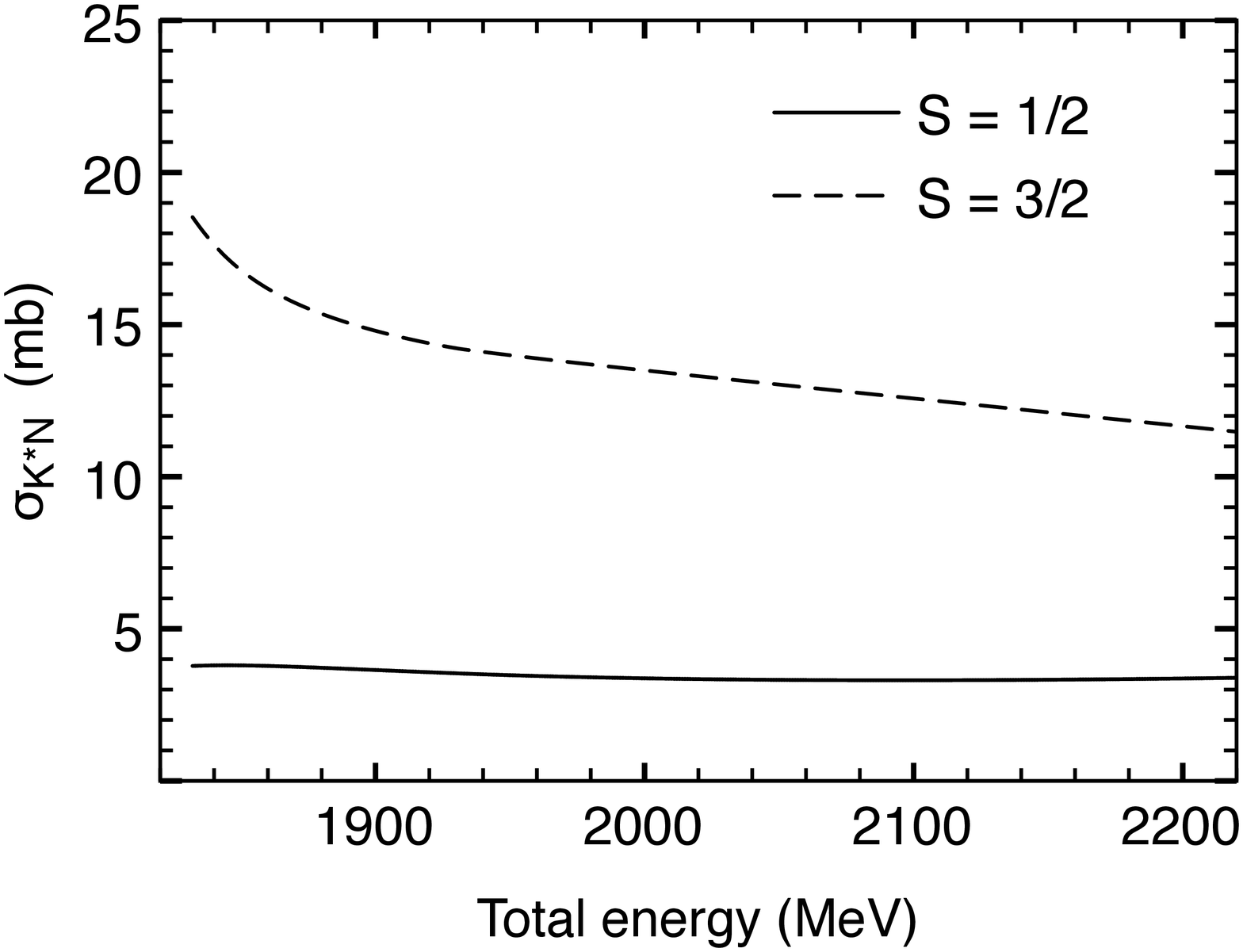}
\includegraphics[height=4.5cm, width=0.6\textwidth]{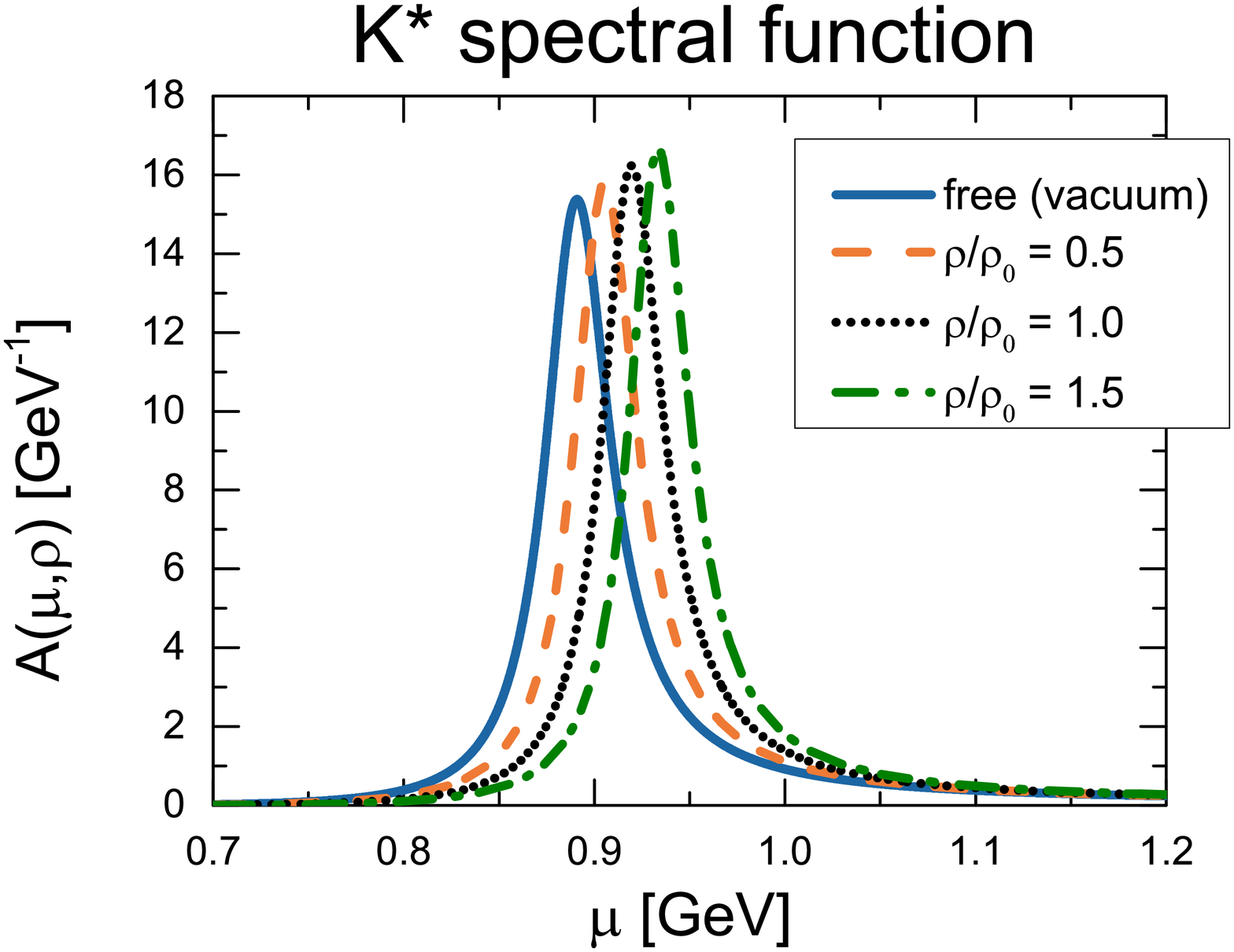}
\caption{Left: Total cross sections  of $K^*N$ in the different spin channels, taken from \cite{Khemchandani:2014ria}. Right: $K^*$ spectral function at zero momentum for different densities, taken from \cite{Ilner:2013ksa}.  \label{fig:kn}}
\end{figure}

With regards to $K^*$, the $KN$ and $K^*N$ interactions are presented using a formalism based on the chiral Lagrangian and the theory of hidden local symmetry. The amplitudes for vector meson-baryon channels are obtained from the $s$-, $t$-, $u$- channel diagrams and a contact interaction, all derived from Lagrangians based on this symmetry. The pseudoscalar meson-baryon interactions are calculated by relying on the Weinberg-Tomozawa theorem. The transition amplitudes between the systems consisting of pseudoscalars and vector mesons are computed  by extending the Kroll-Ruderman term for pion photoproduction replacing the photon by a vector meson.  In addition, the exchange of light hyperon resonances, such as $\Lambda(1405)$ and $\Lambda(1670)$ in the $u$-channel, is included  for the $KN \leftrightarrow KN$,  $K^*N \leftrightarrow K^*N$ and $K^*N \leftrightarrow KN$ processes \cite{Khemchandani:2014ria}.

Once the $s$-wave $KN$ and $K^*N$ interactions are constructed, the Bethe-Salpeter equation is solved in its on-shell factorization form. The subtraction constants required to calculate the loops of the Bethe-Salpeter equation are obtained by fitting the $KN$ amplitudes to the available data for the isospin 0 and 1 $s$-wave phase shifts, separately \cite{Khemchandani:2014ria}.  The  scattering lengths, $a^{I,S}$, at threshold for the $KN$ system in the different isospin sectors  $I$ and $S=1/2$ can be then calculated, i.e., $a^{0,1/2}_{KN} =-0.16$ fm and $a^{1,1/2}_{KN} =-0.29$ fm. The values found by different partial wave analysis groups for the $KN$ scattering lengths range from  $-0.105 \pm 0.01$ fm \cite{prc75} to $-0.23 \pm 0.18$ fm \cite{martin}, for isospin 0, and between $-0.286 \pm 0.06$ fm to $-0.308 \pm 0.003$ fm \cite{prc75}, for the isospin 1 case. Thus, the results are compatible with the available data. Moreover, one can obtain the cross sections  for $K^*N$ (left panel of Fig.~\ref{fig:kn}) and the scattering lengths for different isospin-spin sectors, i.e., $a^{0,1/2}_{K^*N} ({\rm fm})= (0.2,0.03)$, $a^{0,3/2}_{K^*N} ({\rm fm})=(-0.08,0.04)$, $a^{1,1/2}_{K^*N} ({\rm fm})=(0.1,0.0)$ and $a^{1,3/2}_{K^*N} ({\rm fm})=(-0.31,0.03)$.   

As for $K^*$ in nuclear matter, two sources of medium effects  that contribute to the self-energy  in a nuclear medium have been considered in  \cite{Ilner:2013ksa}:  (i) the modification of its dominant decay mode, $K^* \rightarrow K \pi$, induced by medium effects on the light pseudoscalars; and (ii) the quasi-elastic interaction
of the strange meson with nucleons, $K^*N \rightarrow K^*N$ and related absorptive channels with vector mesons. The $K^*$ self-energy has been evaluated straightforwardly in a $T\rho$ approximation, due to the lack of resonant states in the $s=1$ channel \cite{Ilner:2013ksa}.

With all these ingredients, the dynamics of $K^*$ and $\bar K^*$ in HICs has been investigated using the PHSD transport model that
implements in-medium effects on $K^*$ and $\bar K^*$ coming from their production from quark-gluon plasma as well as from the hadronic phase \cite{Ilner:2016xqr}.  The PHSD calculations were performed for Au+Au at $\sqrt{s}_{NN}=$ 200 GeV (STAR/RHIC) and Pb+Pb
$\sqrt{s}_{NN}=$ 2.76 TeV (ALICE/LHC) as well as Au+Au for $\sqrt{s}_{NN}=$~5-60 GeV (CBM/FAIR or
BMN/NICA or BESII/RHIC) by using the off-shell behaviour of $K^*$ and $\bar K^*$. Some of the conclusions are: (i) at LHC/RHIC the main production channel is resonant annihilation of $\pi$ + $K$ ($\bar K$) in the final hadronic phase,
(ii) the baryon densities tested are rather low at LHC/RHIC, so in-medium effects do not play a role, while at lower energies (CBM/BMN/BESII)
the in-medium hadronic effects might be relevant due to the expected longer reaction time and higher densities; and (iii) there are difficulties to extract the in-medium properties of strange vector mesons due to the rescattering and absorption of their decay channels.

\section{Summary}

The properties of strange pseudoscalar and vector mesons as well as strange baryon resonances are reviewed
in hadronic matter. The in-medium modified properties of strange mesons and
baryon resonances are being implemented in transport models to analyze the experimental data from low to high-energy HICs, in order to address some fundamental open question, such as the importance of the hadronic phase (and associated in-medium effects) in their production and propagation.

\section{Acknowledgements}
The author warmly thanks her collaborators Joerg Aichelin, Elena Bratkovskaya, Daniel Cabrera, Kanchan Khemchandani, Alberto Martinez-Torres, Raquel Molina, Eulogio Oset and Angels Ramos for their comments and contributions. The author also acknowledges support from the Ram\'on y Cajal research programme,
FPA2013-43425-P and FPA2016-81114-P Grants from MINECO, NewCompstar COST Action MP1304 and THOR COST Action CA15213.

\end{document}